\documentclass[CRPHYS,Unicode,manuscript]{cedram}
\usepackage{blindtext}
\usepackage{epsfig}
\usepackage{color}
\usepackage{amsmath}
\usepackage{amsfonts}
\usepackage{amssymb}
\usepackage{graphicx}%
\usepackage{url}
\usepackage{esvect}
\usepackage[normalem]{ulem}
\usepackage{xcolor}
\usepackage{xfrac}
\usepackage{dsfont}
\usepackage{bm}
\usepackage{array}
\newcolumntype{C}[1]{>{\centering\let\newline\\\arraybackslash\hspace{0pt}}m{#1}}
\usepackage{footnote}
\usepackage{isomath}
\DeclareMathAlphabet\mathbfcal{OMS}{cmsy}{b}{n}
\usepackage{esint}

\hypersetup{colorlinks,citecolor=blue,filecolor=blue,linkcolor=blue,urlcolor=blue}

 \definecolor{BLACK}{gray}{0}
 \definecolor{WHITE}{gray}{1}
 \definecolor{RED}{rgb}{1,0,0}
 \definecolor{GREEN}{rgb}{0,1,0}
 \definecolor{BLUE}{rgb}{0,0,1}
 \definecolor{CYAN}{cmyk}{1,0,0,0}
 \definecolor{MAGENTA}{cmyk}{0,1,0,0}
 \definecolor{YELLOW}{cmyk}{0,0,1,0}

\title{Quantum Hall and Light Responses in a 2D Topological Semimetal}

\alttitle{R\' eponses de Hall quantique et \` a la lumi\` ere dans un semimetal topologique \` a 2D}

\author{\firstname{Karyn} \lastname{Le Hur}\CDRorcid{0000-0002-3990-4782}\IsCorresp}
\address{CPHT, CNRS, Institut Polytechnique de Paris, Route de Saclay, 91120 Palaiseau, France}
\email[K. Le Hur]{karyn.le-hur@polytechnique.edu}

\author{\firstname{Sariah} \lastname{Al Saati}\CDRorcid{0000-0002-9524-8687}}
\address{CPHT, CNRS, Institut Polytechnique de Paris, Route de Saclay, 91120 Palaiseau, France}
\email[S. Al Saati]{sariah.al-saati@polytechnique.edu}

\begin{abstract}
 We have recently identified a protected topological semimetal in graphene which presents a zero-energy edge mode robust to disorder and interactions. Here, we address the characteristics of this semimetal and show that the $\mathbb{Z}$ topological invariant of the Hall conductivity associated to the lowest energy band can be equivalently measured from the resonant response to circularly polarized light resolved at the Dirac points. The (non-quantized) conductivity responses of the intermediate energy bands, including the Fermi surface, also give rise to a $\mathbb{Z}_2$ invariant. We emphasize on the bulk-edge correspondence as a protected topological half metal, i.e. one spin-population polarized in the plane is in the insulating phase related to the robust edge mode while the other is in the metallic regime. The quantized transport at the edges is equivalent to a $\frac{1}{2}-\frac{1}{2}$ conductance for spin polarizations along $z$ direction. We also build a parallel between the topological Hall response and a pair of half numbers (half Skyrmions) through the light response locally resolved in momentum space and on the sphere.
\end{abstract}

\begin{document}
\maketitle

\section{Introduction}

At the heart of topological aspects in condensed-matter systems is the nontrivial nature of the wavefunction in reciprocal space associated to the band structure of a crystal. Indeed, one of the key theoretical tools underlying the study of topological matter is the notion of topological invariants which have direct consequences on the dynamics of the material. The main topological invariant of interest is the Chern number associated to the Brillouin Zone of the material. Remarkably, this integer Chern number was found to be directly proportional to the Hall conductivity of the material, showing a unique quantized behavior in units of the fundamental constant $e^2/h$ \cite{thouless_quantized_1982,kohmoto_topological_1985}, accessible in quantum transport experiments. This topological number was also found to be equivalently accessible from the response of the material to circularly polarized light \cite{Goldman,Hamburg,PRBLight}. This topological response can be resolved locally within the Brillouin zone through circular light \cite{KLHLight} or using a local microscope i.e. a long transmission line transporting classical light \cite{JulianKaryn}.

Quantum spin Hall insulators are two-dimensional topological insulators which have gained prominence in the last two decades \cite{Dyakonov1971, murakami_dissipationless_2003}. Quantum spin Hall insulators are characterized by the presence of the spin Hall effect, where an electric current induces a transverse spin current due to intrinsic spin-orbit coupling. These spin Hall systems are most often theoretically described by two spin-polarized subsystems, called spin channels, which are related through time-reversal symmetry. A specific example is the case of the Kane-Mele model \cite{kane_quantum_2005,kane_z_2_2005,StephanKaryn,Wu,Wurzburg} in which the spin of the electron is a well defined quantum number labelling distinct energy bands. In this model, each spin channel hosts oppositely directed edge modes, referring then to the quantum spin Hall effect. This description in terms of counter-propagating spin channels is also valid for the Bernevig-Hughes-Zhang model \cite{bernevig_quantum_2006}. The system then hosts a zero total Chern number and is equivalently described with a $\mathbb{Z}_2$ topological spin Chern number leading to spin-Hall currents flowing at the edges of the sample \cite{kane_quantum_2005,Sheng}. Circularly polarized light can measure this topological $\mathbb{Z}_2$ invariant related to the quantum spin Hall effect \cite{KLHLight} and the parity symmetry \cite{FuKane} with a perfect quantization at the Dirac points on the honeycomb lattice associated to the zeros of the Pfaffian \cite{kane_z_2_2005}. 

The classification of topological properties in terms of two distinct Pauli matrices, e.g. related to the two sublattices on the honeycomb lattice and to the two spin polarizations of an electron in the Kane-Mele model, was at the heart of our recent publication \cite{LeHurAlSaati2023}, in which we describe a possible generalization of nontrivial topological properties and robustness towards a specific class of two-dimensional (2D) topological semimetals realizable in graphene. The topological protection of this semimetal was shown by the presence of a zero-energy edge mode polarized along $x$ direction in the energy spectrum robust towards disorder in the presence of interactions that were introduced through a stochastic variational approach \cite{PRBLight,KaneMelestochastic,Kagomestochastic}. 

In this article, we study the bulk topological characteristics of this 2D semimetal in graphene and we present a correspondence between the quantum Hall conductivity and the response to circularly polarized light through the band structure. 
In Sec. \ref{semimetal}, we recall the Hamiltonian and band structure of this system \cite{LeHurAlSaati2023}. We show the validity of the Kubo formula for the topological semimetal, in Sec. \ref{Kubo}. We check that the topological invariant of the lowest-energy band spin-polarized along $x$ direction remains quantized for all the range of parameters in the sense of the quantum Hall response. We also derive the quantum Hall conductivity of the second-energy band and of the nodal ring region. We verify that these contributions are well defined, but these integrals of Berry flux of particles are not quantized reflecting the Fermi surface properties in agreement with the article by Haldane in 2004 \cite{haldane_berry_2004}. Yet we find that we can introduce a quantized $\mathbb{Z}_2$ topological marker to characterize the topological properties associated to the second-energy band and the Fermi surface which is then related to the $\mathbb{Z}$ invariant associated to the lowest-energy band.  The structure of the different edge modes in the band structure crossing the Fermi energy reveals information on both invariants and one zero-energy mode is indeed robust to the existence of bulk modes. An interesting result found in Sec. \ref{light} is that the response to circularly polarized light resolved in frequency, corresponding to a resonance with the Dirac points, can equally reveal this quantized $\mathbb{Z}$ topological marker associated to the lowest-energy band or to the robust edge mode. We also show that the light response for the semimetal is equivalent to a pair of $\frac{1}{2}$ topological numbers (half Skyrmions) from the Dirac points and from the equatorial plane on the sphere. A pair of $\frac{1}{2}$ topological invariants can be detected at the equator from the superposition of the two circularly polarized waves being equivalent to a light source polarized along $x$ direction. This pair of half invariants interestingly also links with the physics of the topological semimetal introduced in the bilayer model related to the model of two entangled spheres \cite{hutchinson_quantum_2021,OneHalfKLH}. This semimetal was understood to host two channels of the Haldane model \cite{haldane_model_1988} as in the Kane-Mele model, revealing a $\mathbb{Z}_2$ symmetry and halved transport at the edges. In the present study in one layer graphene, the bulk-edge correspondence in Sec. \ref{Bulkedge} will lead to the formation of a topological half metal when varying the chemical potential i.e. with one spin polarization showing a quantized conductance and one spin polarization showing a metallic behavior. The quantized transport at the edges along $x$ direction is also equivalent to a $\frac{1}{2}-\frac{1}{2}$ probability for each spin polarization along $z$ direction showing a correspondence with the topological bilayer system \cite{hutchinson_quantum_2021,OneHalfKLH}. 

\section{Protected Topological Semimetal}
\label{semimetal}

The Hamiltonian of the two-dimensional (2D) topological semimetal can be written in tight-binding formalism in reciprocal space as  \cite{LeHurAlSaati2023}
\begin{equation}
    {\mathcal H}=\sum_{\bm k \in BZ} \psi^{\dagger}({\bf k}) {\mathcal H}({\bf k})\psi({\bf k})
    \label{eq:modelsemimetal}
\end{equation}
with 
\begin{align}
    \psi({\bf k})&=(c_{A{\bf k}\uparrow},c_{B{\bf k}\uparrow},c_{A{\bf k}\downarrow},c_{B{\bf k}\downarrow}) \\ \nonumber
    \\
    {\mathcal H}({\bf k}) &=  \bm{d}_{\bm k}\cdot\bm{\sigma}\otimes \mathbb{I} + r\mathbb{I}\otimes s_x,
    \label{eq:Hmodel}
\end{align} where $A$ and $B$ label the two nonequivalent sites of the honeycomb lattice. The Hamiltonian is written in terms of two sets of Pauli matrices: $\bm \sigma$ acting on the Hilbert space $\{|A\rangle;|B\rangle\}$ associated to the two sublattices $A$-$B$ of the honeycomb lattice and $\bf{s}$ acting on $\{|{\uparrow_z}\rangle;|{\downarrow_z}\rangle\}$ corresponding to the two spin polarizations of an electron in the direction perpendicular to the graphene layer. For a spinless system, the sublattice term $\bm{d}_{\bm k}\cdot\bm{\sigma}$ on the honeycomb lattice describes generically a two-band system of energies $\pm \bm{d}_{\bm k}$ where $\bm{d}_{\bm k}$ corresponds to a pseudo vector defined in the Brillouin Zone.  From standard definitions on the honeycomb lattice, the Haldane model \cite{haldane_model_1988} is described by the vector $\bm{d}_{\bm k}$ whose components are given by
\begin{align}
    d_x({\bf k}) &=-t\sum_{i=1}^3 \cos({\bf k}\cdot {\bm{\delta}}_i) \\
    d_y({\bf k}) &=-t\sum_{i=1}^3 \sin({\bf k}\cdot {\bm {\delta}}_i)\\
    {d}_z({\bf k})&=M + 2t_2\sum_{j} \sin({\bf k}\cdot {\bf b}_j),
\end{align}
where $t>0$ represents the hopping amplitude between nearest-neighbours $A$ and $B$, and $\mathbfit{\delta}_i$ represents the three nearest-neighbour vectors linking these sites. For simplicity, we assume here that the second nearest-neighbor hopping term is purely imaginary and equal to $i t_2$ (with $t_2>0$) in the Haldane model. The ${\bf b}_j$ are the three second-nearest neighbour vectors forming a triangle across a unit cell, and the Semenoff mass term $M$ is a parameter describing a modulated potential on the lattice\cite{semenoff_condensed-matter_1984}, which can be implemented e.g. through the presence of a substrate forming a charge density wave. This term induces a different energy shift at the two Dirac points. The specificity of the Haldane model is that the Chern number of the energy bands can be found to be non-zero in the regime where $M < 3\sqrt{3} t_2$ which defines the topological phase. The term $t_2$ is thus the one giving rise to topological properties in graphene, and can be implemented using circularly polarized light \cite{LeHurAlSaati2023, ReviewKLH} as achieved in recent years \cite{Cavalleri}.

\begin{center}
  \begin{figure}[ht]
   \includegraphics[width=0.45\textwidth]{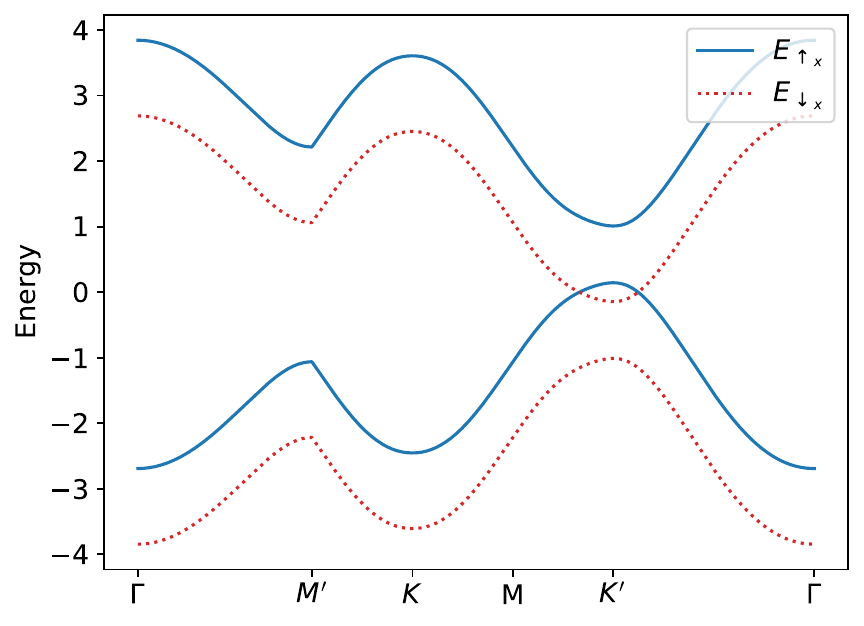}
    \includegraphics[width=0.45\textwidth]{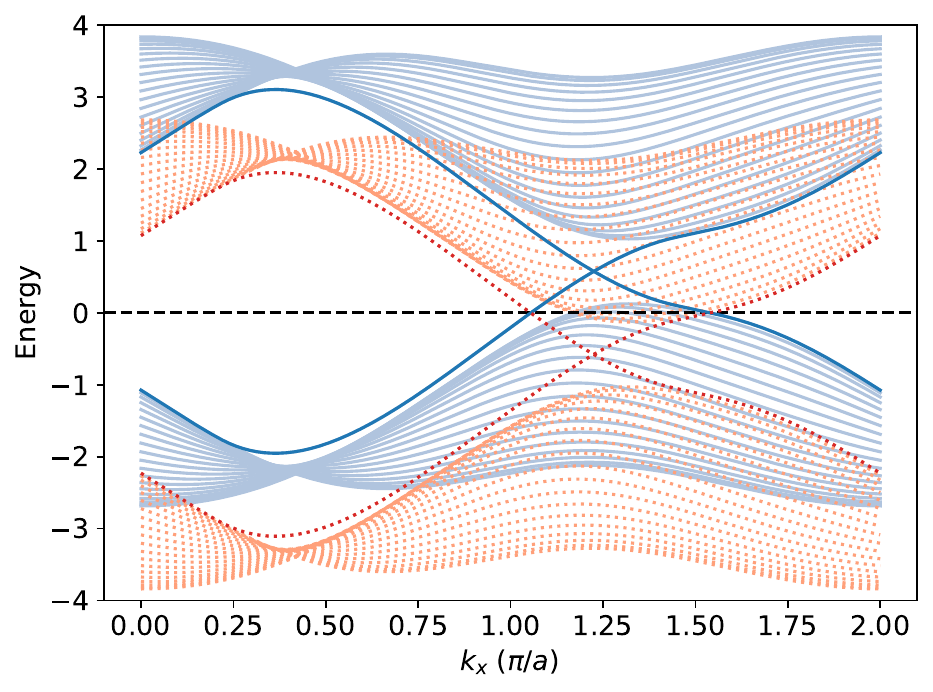}
    \caption{Color Online: (Left) Two-dimensional band structure of the system for a chosen path in the Brillouin zone with the spin polarizations associated to the energy bands shown in distinct colors. 
    (Right) Energy bands showing the structure of the edge modes on a cylinder geometry. The structure of these edge modes crossing the Fermi energy reveals the $\mathbb{Z}_2$ and $\mathbb{Z}$ invariants in Eq. (\ref{quantumnumber}) obtained from the analysis of the quantum Hall responses at half-filling. In momentum space, the band structure with the edge modes can be observed with spin-resolved Angle Resolved Photoemission Spectroscopy (ARPES). In real space, the system shows one robust edge mode in blue (distinct from the bulk modes in red) crossing the Fermi energy. The different responses to circularly polarized light at the $K$ Dirac point will also reveal the quantized quantum Hall conductivity of the lowest-energy band.
    The parameters chosen for all panels are $M = 3\sqrt{3}t/4$, $r = t/\sqrt{3}$, $t_2 = t/3$ with $a=1$.}
   \label{fig:bands}
    \end{figure}
\end{center}

In our study we consider the spin degree of freedom of the electrons. The Hamiltonian given in Eq. (\ref{eq:Hmodel}) includes the term $\bm{d}_{\bm k}\cdot\bm{\sigma}\otimes \mathbb{I}$ where the identity $\mathbb{I}$ acts on the spin degree of freedom. This term thus describes two copies of the same Haldane model, each copy corresponding to a spin projection along the $z$ axis perpendicular to the plane of the lattice. Each band of the Haldane model is this doubly degenerate. To lift this degeneracy, we include a magnetic field, such that its direction is fixed parallel to the layer to induce Zeeman effects. We thus introduce an in-plane Zeeman term described by the $x$ spin-Pauli matrix and given by $r\mathbb{I}\otimes s_x$. This term shifts the two pairs of energy bands by $\pm r$ hence leading to an actual four band system, as shown in Fig. \ref{fig:bands}, with each pair of bands corresponding to an actual spin polarization. Hence, the energy bands of the system are classified according to the spin polarizations along the $x$ direction. In recent years, there have been many  developments of proximity-induced topological phases of matter \cite{Twoplanes,Walter,Leon,Ando} and in particular on graphene/Transition-metal dichalcogenide (TMD) heterostructures \cite{Wang2015, Tiwari2020,ETHZurich} with easily tunable parameters. Therefore, we can anticipate that there will be various equivalent ways to realize this topological semimetal in two dimensions through two sets of Pauli matrices related to the physics of heterostructures which may be equally
realized with cold atoms in optical lattices \cite{hutchinson_quantum_2021,Twoplanes}. It is also important to mention here the observation of the quantum anomalous Hall effect in materials that may be an interesting alternative when including the presence of the staggered potential on the lattice coming e.g. from a substrate
and of the in-plane magnetic field \cite{Chang}. 

This hamiltonian can readily be diagonalized and the Bloch eigenvectors can be expressed in tensorial product of the eigenstates $|\pm\rangle=|\pm_{\bm{d}_{\bm k}}\rangle$ of the operator  $\bm{d}_{\bm k}\cdot\bm{\sigma}$ with eigenvalues $\pm |\bm{d}_{\bm k}|$, tensored with the eigenvectors $|\updownarrow_x\rangle$ of the operator $s_x$ with eigenvalues $\pm 1$. The resulting four possibly overlapping bands are  \begin{align}
    E_{\downarrow_x,-}(\bm k) &= - |\bm{d}_{\bm k}| - r \\
    E_{\downarrow_x,+}(\bm k) &= + |\bm{d}_{\bm k}| - r\\
    E_{\uparrow_x,-}(\bm k) &= - |\bm{d}_{\bm k}| + r \\
    E_{\uparrow_x,+}(\bm k) &= + |\bm{d}_{\bm k}| + r.
\end{align} 

The semimetallic regime is reached when the two middle bands  overlap forming a closed one-dimensional Fermi surface denoted as nodal ring. An important prerequisite to observe the formation of a nodal metallic Fermi liquid is that when solving the equation ${\mathcal H}^2({\bf k})=0=(-r+|{\bf d}|)^2$ this leads to two degenerate solutions corresponding to the two eigenstates $|\uparrow_x\rangle\otimes|-\rangle$ and $|\downarrow_x\rangle\otimes|+\rangle$ such that $2r{\bf d}\cdot\mathbfit{\sigma}\otimes s_x=-2r|{\bf d}|$. The semimetal occurs for the range of parameters $3\sqrt{3}t_2 - M <r < 3\sqrt{3}t_2 + M$ \cite{LeHurAlSaati2023}.
We remind that $3\sqrt{3}t_2$ in the Haldane model refers to $|d_z|$ at the two Dirac points while $d_x,d_y=0$.

The goal of the next Sections is then to link the information on the band structure and edge mode(s) in Fig. \ref{fig:bands} with bulk characteristics such as the quantum Hall conductivity associated to this metallic system and the responses to circularly polarized light
leading to the bulk-edge correspondence in Sec. \ref{Bulkedge}.

\section{Quantum Hall Response and Invariants}
\label{Kubo}

In the linear response regime, the Kubo formula for the Hall conductivity of a 2D electron gas reads \cite{thouless_quantized_1982}
\begin{eqnarray}
        \sigma^{xy} &= & i e^2\hbar  \oiint_{BZ} \frac{d^2\bm{k}}{(2\pi)^2}  \sum_{n \ne m} \frac{f(E_n(\bm{k}))}{(E_n(\bm{k})-E_m(\bm{k}) )^2} \\ \nonumber
        &\times& (\langle n \bm{k} |  v^x  |m \bm{k}\rangle \langle m \bm{k}|v^y |n \bm{k}\rangle - \langle n \bm{k} |  v^y  |m \bm{k}\rangle \langle m \bm{k}|v^x |n \bm{k}\rangle)
        \label{eq:KuboCond}
    \end{eqnarray} 
where the indices $n = \updownarrow_x, \pm$ and $m = \updownarrow_x^\prime, \pm^\prime$ run over the four bands of the system, and $f(x) = \theta(x)$ is the Fermi-Dirac distribution function being equal to the Heaviside step function at zero temperature.

One may question the applicability of this formula in the semimetallic regime for which bands $\downarrow_x,+$ and $\uparrow_x,-$ cross zero energy such that $E_n(\bm{k})-E_m(\bm{k})=0$.
Here, a divergence does not occur because the current operator 
\begin{align}
    v^\alpha = \dfrac{1}{\hbar} \partial_{k_\alpha}\mathcal{H} \otimes \mathbb{I},
\end{align} shows the important property $\langle \uparrow_x,-| v^\alpha | \downarrow_x,+ \rangle = 0$. Eq. (\ref{eq:KuboCond}) is thus well defined and we can rewrite this expression in terms of the integral of the Berry curvature on the Brillouin zone and all occupied states 
\begin{align}
    \sigma^{xy}  &= \dfrac{e^2}{h}  \sum_{n} \oiint_{BZ}\dfrac{\mathrm{d}^2\bm{k}}{2\pi} f(E_n(\bm{k}))  \mathcal{B}_{n\bm{k}}
    \label{eq:KuboChanged}
\end{align}
where the Berry curvature of band $n$ is defined as \cite{Berry}
\begin{align}
    \mathcal{B}_{n\bm{k}} =  i\left( \partial_{k_x} \langle n |\right) \partial_{k_y} | n  \rangle   - i\left( \partial_{k_y} \langle n |\right)  \partial_{k_x} | n  \rangle.
\end{align} 

In the semimetallic regime, the Fermi-Dirac distribution with the term $f(E_n(\bm{k})) $ appearing in Eq. (\ref{eq:KuboChanged}) plays a major role as it constrains the integration to be performed only partially over the Brillouin zone for the two middle bands $n = \downarrow_x,+$ and $n=\uparrow_x,-$, corresponding to two partially filled bands, while the integration involves the entire Brillouin zone for $n = \downarrow_x,-$ whose band is entirely filled. Correspondingly, the band $n = \uparrow_x,+$ gives no contribution to the resulting conductivity as it is completely empty. Indeed, the Fermi level is set at $\mu = 0$ and we consider the sample to be at half-filling. This expression gives \begin{align}
    \sigma^{xy} &= \dfrac{e^2}{h} \left( C_{\downarrow_x,-} +\widehat{C}_{\uparrow_x,-} + \widetilde{C}_{\downarrow_x,+}  \right)
    \label{eq:Cs}
\end{align} for which we have defined (hat and tilde symbols) \begin{align}
    C_n &=  \dfrac{1}{2\pi} \iint_{BZ} \mathrm{d}^2\bm{k}  {\mathcal{B}}_{n\bm{k}} \\
    \widehat{C}_{n} &= \dfrac{1}{2\pi} \iint_{\overline{NoR}} \mathrm{d}^2\bm{k} {\mathcal{B}}_{n\bm{k}} \\
    \widetilde{C}_{n} &= \dfrac{1}{2\pi} \iint_{NoR} \mathrm{d}^2\bm{k}  {\mathcal{B}}_{n\bm{k}}.
\end{align} 
For the regular Chern number $C_n$ (with $n=\downarrow_x,-$ associated to the lowest filled energy band), the integral is performed over $BZ$ the entire Brillouin zone, while it refers to the region $\overline{NoR}$ outside the nodal ring for $\widehat{C}_n$ 
(associated to the partially filled band $n=\uparrow_x,-$) and to the region $NoR$ inside the nodal ring for $\widetilde{C}_n$ (with $n=\downarrow_x,+$). The first integral running on the whole Brillouin zone, associated to the lowest filled energy band, is characterized through an integer topological number similarly as in the Haldane model, assuming that $3\sqrt{3}t_2>M$.

It is useful to formulate a correspondence from the Brillouin zone to 2-sphere. The $\bm d_{\bm k}$ vector appearing in the Hamiltonian of Eq. (\ref{eq:Hmodel}) leads to the following well-defined  mapping from the Brillouin Zone to 2-sphere  \cite{Cayssol}
\begin{align}
    \Phi : BZ &\rightarrow S^2 \nonumber\\
    \bm k &\mapsto (\theta_{\bm k},\varphi_{\bm k})
\end{align} where $(\theta_{\bm k},\varphi_{\bm k})$ are the two uniquely defined angles on $S^2 = \left[0, \pi\right[ \times \left[0, 2\pi \right[$ such that \begin{align}
    \bm d_{\bm k} = |\bm d_{\bm k}| \begin{pmatrix}
        \sin \theta_{\bm k} \cos \varphi_{\bm k}  \\
        \sin \theta_{\bm k}  \sin \varphi_{\bm k} \\
        \cos \theta_{\bm k}
    \end{pmatrix}.
\end{align}
Here, $\varphi_{\bm k}$ and $\theta_{\bm k}$  are the azimuthal and polar angles of the three-dimensional vector $\bm d_{\bm k}$. This map gives us \begin{align}
    \mathcal{B}_{\downarrow_x,-,\bm{k}} &= -\dfrac{1}{2} \sin \theta_{\bm k} \left(\partial_{k_x}\theta_{\bm k} \partial_{k_y}\varphi_{\bm k} - \partial_{k_x}\varphi_{\bm k} \partial_{k_y}\theta_{\bm k} \right) \\
    \mathcal{B}_{\uparrow_x,+,\bm{k}} &= +\dfrac{1}{2} \sin \theta_{\bm k} \left(\partial_{k_x}\theta_{\bm k} \partial_{k_y}\varphi_{\bm k} - \partial_{k_x}\varphi_{\bm k} \partial_{k_y}\theta_{\bm k} \right)\\
    \mathcal{B}_{\downarrow_x,\pm,\bm{k}} &= \mathcal{B}_{\uparrow_x,\pm,\bm{k}}.
\end{align} We can integrate by change of variables $\bm k \rightarrow (\theta,\varphi)$, where we recognize $\partial_{k_x}\theta_{\bm k} \partial_{k_y}\varphi_{\bm k} - \partial_{k_x}\varphi_{\bm k} \partial_{k_y}\theta_{\bm k}$ as the Jacobian of the transformation. The Fermi surface formed by the nodal ring can be expressed to a good approximation in these variables by the contour $\theta = \theta_c, \varphi \in \left[0, 2\pi\right]$ for a given $\theta_c$ fixed between $0$ and $\pi$ depending on the value of $r$, which defines the position of the nodal ring in the Brillouin zone.

We get by change of variables  \begin{align}
    \widetilde{C}_{\downarrow_x,+} &= \left(\dfrac{1}{2\pi}\int_{0}^{2\pi}\mathrm{d}\varphi \right) \int_{\theta_c}^\pi \mathrm{d}\theta  (+\dfrac{1}{2}) \sin \theta \nonumber \\  
     &= \dfrac{1}{2}\left[ \cos \theta_c + 1\right]  \\
    \widehat{C}_{\uparrow_x,-} &= \left(\dfrac{1}{2\pi}\int_{0}^{2\pi}\mathrm{d}\varphi \right) \int_{0}^{\theta_c} \mathrm{d}\theta  (-\dfrac{1}{2}) \sin \theta  \nonumber \\
     &= \dfrac{1}{2}\left( \cos \theta_c - 1\right) \\
    C_{\downarrow_x,-} &= \left(\dfrac{1}{2\pi}\int_{0}^{2\pi}\mathrm{d}\varphi \right) \int_{0}^{\pi} \mathrm{d}\theta  (-\dfrac{1}{2}) \sin \theta \nonumber \\
     &= -1,
\end{align} resulting in the Hall conductivity \begin{align}
    \sigma^{xy} &= \sigma^{xy}_{\downarrow_x,-} +\widehat{\sigma}^{xy}_{\uparrow_x,-} + \widetilde{\sigma}^{xy}_{\downarrow_x, +} \nonumber \\
    &= \dfrac{e^2}{h}\left( \cos \theta_c - 1\right).
    \label{eq:Conductivity}
\end{align}
This expression interpolates between $\sigma^{xy} = -2\dfrac{e^2}{h}$ for the topological insulating phase when $r < 3\sqrt{3}t_2 - M$ for which $\theta_c = \pi$ all the way to $\sigma^{xy} = 0$ for the trivial insulating phase when $r > 3\sqrt{3}t_2 + M$ for which $\theta_c = 0$. The nodal ring equivalently navigates from the vicinity of $K'$ towards $K$ when increasing  $r$ within the semimetal region such that $3\sqrt{3}t_2 - M< r < 3\sqrt{3}t_2 + M$. However, the quantized invariant for the $C_{\downarrow_x,-}$ term associated to the lowest-energy band characterizes the presence and robustness of the zero-energy edge mode and can therefore represent a good marker for this topological system. Now, regarding the other two terms traducing the Berry flux accumulated by particles moving around the Fermi surface \cite{haldane_berry_2004}, we can in fact also introduce a quantized $\mathbb{Z}_2$ invariant, valid for all range of parameters within the semimetal metal phase, such that 
\begin{equation}
\label{quantumnumber}
\hat{C}_{\uparrow_x,-} - \tilde{C}_{\downarrow_x,+} = C_{\downarrow_x,-}.
\end{equation}
In this sense, we can introduce two invariants to characterize this energy band structure i.e. the $\mathbb{Z}$ quantized invariant related to the lowest energy band and a $\mathbb{Z}_2$ index formed through the second-energy band and the nodal ring region or Fermi surface.
We observe that this topological marker revealing both the $\mathbb{Z}_2$ character associated to the second-energy band and the $\mathbb{Z}$ invariant from the lowest-energy band is revealed in the structure of the edge modes associated to the different energy bands
\ref{fig:bands}.

Here, we emphasize that the introduced topological invariants are yet applicable including interactions between electrons. For instance, we can introduce a Hubbard interaction $H_U=U\sum_i \hat{n}_{i\uparrow}\hat{n}_{i\downarrow}$ with the densities
associated to each spin polarization along $z$ direction at a site $i$ of the lattice. Through the stochastic variational approach that we recently introduced for the interacting Haldane \cite{PRBLight} and Kane-Mele \cite{KaneMelestochastic} models we observe that the $U$ term produces in particular spin channels along the three directions x,y,z, \{$\phi_x$, $\phi_y$, $\phi_z$\}, respecting the $SU(2)$ invariance of the system when $r=0$. The mean-field variational solution then leads to $\phi_r = -\frac{1}{2}\langle c^{\dagger}_i \bm{s} c_i\rangle$. This approach includes the different channels in momentum space directly in the local theory $\mathcal{H}({\bf k})$ when addressing the uniform and zero-frequency limit in the action to study ground-state properties. We have also shown a good quantitative agreement of this method with the best numerical methods already at a mean-field level where the Mott transition can be found analytically in one equation for the Haldane and Kane-Mele models \cite{PRBLight,KaneMelestochastic}. When $r=0$, this approach shows the solution $\phi_x=\phi_z=\phi_y=0$ at weak (moderate) interactions. Including an in-plane magnetic field will fix the difference in momentum space between the number of spins-$\downarrow_x$ and the number of spins-$\uparrow_x$ to be equal to two times the area of the nodal ring $\sim 4\pi r_c^-(r - r_c^-)$ with $r_c^- = (3\sqrt{3}t_2-M)$ such that $\phi_x$ scales as $2\pi \frac{r_c^-}{N}(r - r_c^-)$ with $N$ being the number of sites in the lattice whereas $\phi_z=0=\phi_y$. Therefore, at weak interactions, the spin channel $\phi_x$ slightly dresses the effect of the in-plane magnetic field.  From this observation, we justify the robustness of the topological quantum Hall response and of the $\mathbb{Z}_2$ topological invariant for weak (to moderate) interactions. 

Below, we show that the quantized topological response of the lowest band associated to $C_{\downarrow_x,-}$ can be measured through circularly polarized light which then reveals the presence of the robust edge mode. We will also show that this response locally measured
from the resonance at the Dirac points is mathematically equivalent to measure the quantum Hall conductivity with $\theta_c=\frac{\pi}{2}$ on a sphere i.e. revealing a half-Skyrmion, linking with the physics of the bilayer model \cite{hutchinson_quantum_2021,OneHalfKLH}.
Fixing the polar angle in the equatorial plane on a sphere, we find that the light can also reveal a pair of half invariants corresponding then to the response to a linear polarization along $x$ direction i.e. corresponding to the superposition of left- and right-handed circular waves.

\section{Topological Markers from Circularly Polarized Light}
\label{light}

We measure the topological response from a circularly polarized light field in the plane \cite{Goldman} localized resolved in frequency to reach the Dirac point(s) \cite{PRBLight,ReviewKLH,KLHLight}.
In Appendix \ref{lightmatter}, we introduce the definitions of the light-matter interaction on the honeycomb lattice and remind the main steps of our approach to measure the topological properties locally in momentum space at the two Dirac points for the two-bands Haldane model.
The goal in this Section is then to generalize this approach for the topological semimetal introducing local topological markers in momentum space from the responses to these classical electromagnetic waves, e.g. through Eq. (\ref{response}) below.

For the topological semimetal, within the Dirac approximation for the $d_x({\bf k})$ and $d_y({\bf k})$ terms, the Hamiltonian takes the form
\begin{equation}
{\mathcal H}({\bf p}) = \left(\hbar v_F (p_x \sigma_x + \zeta p_y \sigma_y) + (\zeta d_z +M) \sigma_z \right)\otimes \mathbb{I} + r \mathbb{I}\otimes s_x,
\end{equation}
and $p_x$ and $p_y$ are wave-vector components measuring deviations from each Dirac point. Within the choice of Brillouin zone, as in Refs. \cite{PRBLight,ReviewKLH,KLHLight}, 
the two Dirac points are related through the symmetry $p_x\rightarrow p_x$ and $p_y\rightarrow -p_y$ such that $\zeta=\pm$ at the $K$ and $K'$ Dirac points respectively. 

To generalize the forms of the photo-induced currents for the topological semimetal with four bands, we can proceed as in Appendix \ref{lightmatter} for the honeycomb lattice and the two-bands Haldane model and introduce the generalized definitions
$\sigma_x\otimes\mathbb{I} = \frac{1}{\hbar v_F}\frac{\partial {\mathcal H}}{\partial p_x} \otimes \mathbb{I}$ and $\sigma_y \otimes \mathbb{I} = \frac{1}{\hbar v_F}\frac{\partial {\mathcal H}}{\partial (\zeta p_y)} \otimes \mathbb{I}$.
The photo-induced current operator then takes the form
\begin{equation}
\hat{J}_{\zeta}(t) = \left(\frac{1}{2\hbar}A_0 e^{-i\omega t} \left( \frac{\partial {\mathcal H}}{\hbar\partial (\zeta p_y)} +\zeta i  \frac{\partial {\mathcal H}}{\hbar\partial p_x}\right) +h.c.\right)\otimes \mathbb{I}.
\end{equation}
This equation is then a generalization of Eq. (\ref{Jzeta}) where we introduce the vector potentials associated to each light polarization in Eq. (\ref{vectorpotential}), $A_0$ being the amplitude of the light field for each polarization.
Due to the pseudo-spin structure acting on the sub-lattice space, to evaluate the variation of the charge density in time and therefore the current density from the continuity equation, it is useful to introduce the projectors on each sub-lattice $\hat{P}_a = \frac{1}{2}(\mathbb{I}+\sigma_z)$ and 
$\hat{P}_b = \frac{1}{2}(\mathbb{I}-\sigma_z)$ \cite{PRBLight,ReviewKLH,KLHLight}. In this way, the $d_z$ term commutes with the charge density operator in momentum space and does not modify the structure of the current operator compared to the graphene Hamiltonian through the Heisenberg equations of motion. At the Dirac points, to have a photo-induced current this requires to flip the pseudo-spin operator as the lower and upper bands are related through distinct quantum numbers $|+\rangle$ and $|-\rangle$. Such a form of current operator in Eq. (\ref{Jzeta}) can also be found within a general approach of dipole theory \cite{Goldman}. The topological structure of eigenstates will then intervene when performing the linear response theory in $A_0$. This is also correct for the four bands model, i.e. the topological semimetal, since light does not modify the spin quantum number therefore, e.g. at the $K$ Dirac point, this will only mediate transitions between the pair of bands $n=\downarrow_x,-$ and $m=\downarrow_x,+$ and similarly between the pair of bands $n^\prime=\uparrow_x,-$ and band $m^\prime=\uparrow_x,+$. Since the same light polarization (the right-handed light) will couple these pairs of bands, we can introduce the same symbol $\zeta$ specifying a Dirac point and a light polarization. This is the important prerequisite to observe circular dichroism in these topological band structures, one light polarization interacting specifically with one Dirac point. It is also important to add that since we have an identity matrix in the spin sector, this refers to the current of a spin-$\uparrow_z$ or spin-$\downarrow_z$ particle equally.  We verify below through justifications from the band structure that since we have an identity operator in the spin space for the current it gives equal currents for a spin-$\uparrow_x$ or spin-$\downarrow_x$ particle. Since we measure the response to the electric field, the magnetic term $r$ will not modify the form of the current operator, but it will then again intervene in the important details hidden in the band structure when performing the linear response theory in $A_0$ e.g. through the energy conservation.

An important point is then to incorporate the energy conservation within the four-energy bands model describing the topological semimetal. It is useful to write down the light-matter interaction $\delta H_{\pm} = A_0 e^{\pm i\omega t}\sigma^+\otimes\mathbb{I}$ \cite{ReviewKLH,KLHLight} within the four energy bands. The indices $\pm$ refer respectively to a right-handed and left-handed light polarizations. Since the light-matter interaction couples bands with the same spin quantum number, this approach verifies indeed that at the $K$-point this will only mediate transitions between bands $n=\downarrow_x,-$ and band $m=\downarrow_x,+$ and between bands $n^\prime=\uparrow_x,-$ and band $m^\prime=\uparrow_x,+$. Interestingly, it is possible to find a resonance with both pairs of bands simultaneously fixing the light frequency such that $\hbar\omega=2|d_z(K)|=6\sqrt{3}t_2+2M$ e.g. for the right-handed wave (+). Due to the structure of eigenstates in the valley around the $K$ Dirac point, in this way only one of the two light polarizations will be able to mediate inter-band transitions similarly to the topological two-bands model. Thus, due to the identification between eigenstates $|\uparrow_z\rangle=\frac{1}{\sqrt{2}}(|\uparrow_x\rangle +|\downarrow_x\rangle)$ and $|\downarrow_z\rangle=\frac{1}{\sqrt{2}}(|\uparrow_x\rangle -|\downarrow_x\rangle)$,  we deduce that the inter-band transitions for one spin polarization $\uparrow_z$ or $\downarrow_z$ will then correspond effectively to a half of the sum of the transition rates associated to the two pairs of bands $(n,m)$ and $(n^\prime, m^\prime)$ respectively, which are in fact equal. Therefore, for the topological semimetal, the response at the $K$ point for the $+$ light polarization is identical to $\Gamma_+(\omega)$ in Eq. (\ref{zeta}) for each spin polarization $|\uparrow_z\rangle$, $|\downarrow_z\rangle$ or for each pair of bands $(n,m)$ and $(n^\prime, m^\prime)$ corresponding to spin polarizations $|\downarrow_x\rangle$ and $|\uparrow_x\rangle$ respectively. 

At the $K'$ point, the two occupied bands ($n$ and $m$) at zero temperature have the same spin polarization  whereas the two empty bands ($n'$ and $m'$) have now a distinct spin polarization. Therefore, to mediate inter-band transitions towards empty states in the quantum limit, this would require to also flip the spin polarization of an electron. Therefore, in the presence of the light field, due to the two-particles ground-state structure at the $K'$ point, this results in $\Gamma_- = 0$ \cite{ReviewKLH}. Hence, if we define circular dichroism for a pair of energy bands, e.g. related to the first (second) and third (fourth) energy bands at the two Dirac points, we also have $\Gamma_-(\omega)=0$ for these pairs at finite temperature due to the left-handed polarized electromagnetic wave. 

It is then useful to analyze the information in $\Gamma_+$ at the $K$ point from Eq. (\ref{zeta}) through the Berry curvature $F_{p_x p_y}$ mapped onto the 2-sphere \cite{ReviewKLH} where the $K$ Dirac point corresponds to north pole and $K'$ to south pole along the polar angle direction. 
There is then a simple relation between the Berry curvature in the plane and the pseudo-spin variable \cite{KLHLight}
\begin{equation}
\label{lightberry}
F_{p_x p_y} = \frac{2(\hbar v_F)^2}{|E_u-E_l|^2}\cos\theta = \frac{2(\hbar v_F)^2}{|E_u-E_l|^2} \langle \sigma_z(\theta)\rangle
\end{equation}
where for the topological semimetal, at the $K$ point, $E_u-E_l=E_m-E_n=E_{m^\prime}-E_{n^\prime}$. See Appendix \ref{Berry} for a justification of the map between momentum space and sphere in this case from the Dirac approximation for the kinetic term
which allows us to study the effect of circular motion in the equatorial plane of the sphere and also around each Dirac point in the Brillouin zone. This form of Berry curvature on the lattice can be verified from the Dirac points within the linear approximation for the kinetic term \cite{ReviewKLH,Ryu} and it reproduces exactly the quantum Hall response. Therefore, for the topological semimetal, at the $K$ point for a spin polarization $\uparrow_z$, $\downarrow_z$ or a pair of bands $(n,m)$ and $(n^\prime, m^\prime)$ we have
\begin{equation}
\left|\frac{\Gamma_+}{2}\right| =  \frac{2\pi}{\hbar}A_0^2\frac{1}{2}\langle \sigma_z(0)\rangle.
\end{equation}
Since at the $K'$ point, $\Gamma_-=0$ then this is equivalent to
\begin{equation}
\left|\frac{\Gamma_+ - \Gamma_-}{2}\right| =  \frac{2\pi}{\hbar}A_0^2\frac{1}{2}\left| \langle \sigma_z(0)\rangle \right|.
\end{equation}
Here, $\sigma_z$ measures the relative density on the two sublattices of the honeycomb lattice resolved at the $K$ Dirac point. In comparison, for the Haldane model, we obtain
\begin{eqnarray}
\left|\frac{\Gamma_+ - \Gamma_-}{2}\right| &=&  \frac{2\pi}{\hbar}A_0^2\frac{1}{2}\left|(\langle \sigma_z(0)\rangle - \langle \sigma_z(\pi)\rangle)\right| \nonumber \\ 
&=& \frac{2\pi}{\hbar}A_0^2|\langle \sigma_z(0)\rangle|.
\end{eqnarray}
The topological number of the Haldane model on the 2-sphere takes the equivalent form \cite{Dynamo,ReviewKLH}
\begin{equation}
C = \frac{1}{2}(\langle \sigma_z(0)\rangle - \langle \sigma_z(\pi)\rangle).
\end{equation}
 It is interesting to observe that within this map from momentum space to surface of the sphere, the quantum Hall response can also be revealed from the two Dirac points (see Appendix \ref{Berry}) \cite{ReviewKLH}.
 
A peculiarity of this topological semimetal is that $\langle \sigma_z(\pi)\rangle=0$ as a result of the formation of the nodal ring semimetal around $K'$ which is another way to justify why $\Gamma_-=0$ in this case.
The response for one-spin polarization is now {\it halved} compared to one topological plane as a result of $\Gamma_-=0$.  The topological semimetal in the bilayer model \cite{hutchinson_quantum_2021,OneHalfKLH} can be characterized in terms of a pair of one-half local topological markers and $\pi$-Berry phases \cite{OneHalfKLH} related to the two-spheres' model associated to spin polarizations $\uparrow_z,\downarrow_z$ \cite{hutchinson_quantum_2021}. In the one-layer graphene realization, the $\pi$ Berry phase is associated equally to the phase acquired by an electron with spin polarization $\uparrow_x$ in band $2$ and by an electron with spin polarization $\downarrow_x$ in band $1$, when circling around the $K$ Dirac. A pair of $\pi$-Berry phases is also analogous to a pair of $\pi$ winding numbers. A 
$\frac{1}{2}$ light response is also predicted to occur for the $\nu=\frac{1}{2}$-fractional quantum Hall effect \cite{CecileNathan}. 
 If we sum the information on these two spin polarizations (or equally $\uparrow_z,\downarrow_z$) i.e. on the two pairs of bands $(n,m)$ and $(n^\prime, m^\prime)$, we have
\begin{equation}
\label{response}
\sum_{j=\uparrow;\downarrow}\left|\frac{\Gamma_+ - \Gamma_-}{2}\right| =  \frac{2\pi}{\hbar}A_0^2\cos(\theta=0).
\end{equation}
Now, we can establish a precise relation with the quantum Hall response of the lowest energy band in Eq. (\ref{eq:Conductivity}) through
\begin{equation}
C_{\downarrow_x,-} = \frac{1}{2}(\cos(\pi)-\cos(0)) = - \cos(0) = -1.
\end{equation}
Therefore, we obtain a relation between light and topological response associated to the lowest-energy band
\begin{equation}
\label{lightformula}
\sum_{j=\uparrow,\downarrow}\left|\frac{\Gamma_+ - \Gamma_-}{2}\right| =  \frac{2\pi}{\hbar}A_0^2 |C_{\downarrow_x,-}|.
\end{equation}
When ${r} < 3\sqrt{3}t_2 - M$, the left-handed polarization $(-)$ will also mediate transitions between the pair of bands $(n^\prime, m^\prime)$ at the $K'$ Dirac point measuring twice the topological invariant of the lowest band, in agreement
with the quantum Hall conductivity. For ${r}>3\sqrt{3}t_2 - M$, one pair of bands will  interact with one distinct light polarization at each Dirac point. In this case, if we reproduce the steps of the calculation above we now find
that $\Gamma_+ - \Gamma_- =0$ at each Dirac point. In Appendix \ref{lightmatter} and Eq. (\ref{lightequation}), the two pairs $(n,m)$ and $(n^\prime, m^\prime)$ are now characterized with different signs of $A_y(t)$ and inverted lower and upper energy eigenstates, for the same Dirac point.

The light response at the $K$ and $K'$ points for the topological Haldane model reveals the same information as the light response and therefore the quantum Hall response coming from a summation on the whole Brillouin zone \cite{ReviewKLH}; see also Appendix \ref{lightmatter}.
This identification necessitates that the quantum Hall responses and light responses are continuous and differentiable everywhere. For the topological semimetal, this requires that the intermediate bands do not alter the result from the lowest-energy band in Sec. \ref{Kubo}. In this way, the light response from the Dirac points is identical to the quantum Hall response of the semimetal only if $\theta_c=\frac{\pi}{2}$. For $\theta_c=\frac{\pi}{2}$, we also observe that the quantum Hall response coming from $\uparrow_x$ particles in the second-energy band takes the form  $\frac{e^2}{h}\widehat{C}_{\uparrow_x,-}=-\frac{1}{2}\frac{e^2}{h}$ \cite{OneHalfKLH}. Similarly, if we now sum the topological number associated to the $\downarrow_x$ particles in the two energy bands, this also measures $\frac{e^2}{h}(\widetilde{C}_{\downarrow_x,+} + C_{\downarrow_x,-})=-\frac{1}{2}\frac{e^2}{h}$ and gives a similar interpretation to the other half-topological number. The light approach then equivalently measures the topological response $\sum_{j=\uparrow,\downarrow} C_{\alpha}$ with $|C_{\alpha}|=\frac{1}{2}$, in relation with the bilayer model 
and a local formulation of a one-half topological invariant for the semimetal \cite{hutchinson_quantum_2021,LeHurAlSaati2023,OneHalfKLH}.  

Interestingly, this is also possible to reproduce the half quantization from the light responses if we fix $\theta_c=\frac{\pi}{2}$ when measuring the interband transition probabilities. See Appendix \ref{Berry}.

\section{Bulk-Edge Correspondance}
\label{Bulkedge}

The topological semimetal of interest in this paper can also be understood as a superposition of two Haldane models associated here to each spin polarization towards the axis of the in-plane magnetic field corresponding to the $r \mathbb{I}\otimes s_x$ term in the Hamiltonian of Eq. (\ref{eq:Hmodel}). This magnetic field of small amplitude polarizes the system into two subsystems which are independent of each other as long as the $\mathbb{Z}_2$ symmetry of the Hamiltonian is preserved.
\begin{center}
    \begin{figure}[ht]
    \includegraphics[width=0.45\textwidth]{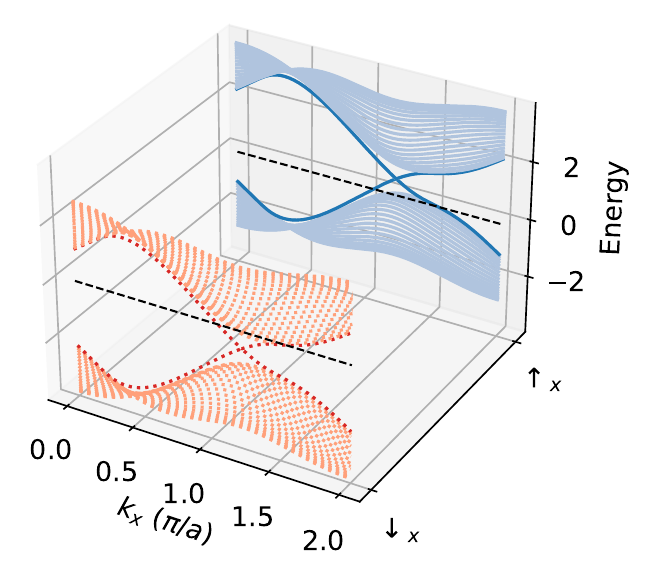}
    \caption{Color online: One-dimensional energy bands of the topological semimetal showing the two underlying band structures of opposite spin polarization, with the spin polarization shown in the third direction. The parameters chosen for all panels are $M = 3\sqrt{3}t/4$, $r = t/\sqrt{3}$, $t_2 = t/3$ with $a=1$.}
      \label{fig:bandscyl}
    \end{figure}
\end{center}
As shown in Fig. \ref{fig:bandscyl}, these two sub copies of the Haldane model are of the same topological nature in the sense that the lower bands have the same Berry curvature and hence the same Chern number, but the bands of the two sub systems are not filled evenly, as shown in Fig. \ref{fig:bandscyl}. The second band of the $\downarrow_x$ population is partially filled while the first band of the $\uparrow_x$ population is partially filled symmetrically to the $\downarrow_x$ population. As long as the  $\mathbb{Z}_2$ symmetry of the system is preserved, the two subsystems are well defined and the lower band of the $\uparrow_x$ population does not mix to the upper band  of the $\downarrow_x$ population, and no gap is opened in the band structure. 

The topological nature of this semimetal comes from the fact that each of the spin subsystem acts as if its chemical potential was shifted by $\pm r$ from its half-filling value, which leads for the bands to be either emptied or filled in a symmetric way between the two subsystems. This does not change however the topological nature of the bands themselves which keep their Berry curvature and their resulting Chern number. The topological nature of the bands is well defined here in terms of Chern number, but the resulting band structure is gapless when taking the two spin populations together. ARPES appears indeed as the most simple and decisive way to probe this semimetal. This should allow us to retrieve the band structure of each spin population independently, and thus to characterize the topological nature of each spin population on its own. This approach should also allow for example to reproduce the topological nature of the spin Hall phase for the case of the Kane-Mele model. 

Another way to probe this semimetal is using spin-polarized transport at chemical potential $\mu$ which satisfies $|\mu \pm r|< |M - 3\sqrt{3} t_2|$. Indeed, the spin polarization of the system in the $x$ direction is a well defined quantum number and in such a regime for the chemical potential, the mesoscopic Hall conductance of one spin population is quantized due to the unique presence of spin-polarized edge modes while the conductance of the other population is non quantized due to the presence of spin-polarized bulk modes at this regime of chemical potential.  If we consider the graphic representations of the energy bands in Fig. \ref{fig:bandscyl}, the system can thus be seen in this regime of chemical potential as a topological half metal: one spin population is in the topological insulating phase while the other is in a metallic regime. We also emphasize here that the presence of a quantized conductance at the edges for one spin-population polarized along $x$ direction is also equivalent to a $\frac{1}{2}-\frac{1}{2}$ conductance for the two spin polarizations along $z$ direction which is precisely at the heart of the bulk-edge correspondence interpretation for the bilayer model \cite{hutchinson_quantum_2021,OneHalfKLH} and the $\mathbb{Z}_2$ symmetry associated to the two spin polarizations along $z$ direction.
 
\section{Conclusion}
\label{Conclusion}

To summarize, the quantum Hall conductivity of the recently identified two-dimensional topological nodal ring semimetal on the honeycomb lattice in one-layer graphene reveals
a $\mathbb{Z}$ topological invariant $C_{\downarrow_x,-}$ associated to the lowest-energy band and a $\mathbb{Z}_2$ spin Chern number $\hat{C}_{\uparrow_x,-} - \tilde{C}_{\downarrow_x,+}$ for the
partially filled bands.  The topological semimetal phase, associated to  the occurrence of a gap closing region at the Fermi energy due to crossing between two bands, is also characterized through a $\mathbb{Z}_2$ spin Chern number
in addition to the quantized quantum Hall response. We evaluated the response to circularly polarized light, which is an equivalent marker of the topological response and of the quantum Hall conductivity for the Haldane model, and showed how this can yet reveal the quantum Hall conductivity for the lowest-energy band of the topological semimetal from the frequency resonance corresponding to the energy gap at the Dirac points. 

We have then justified that this system is a topological half metal, i.e. one spin population along $x$ direction being in the topological insulating phase while the other is in a metallic regime resulting in a conductance at the edges quantized for one spin-polarization only. The  quantized edge mode then has a 1/2 probability to be in the $\uparrow_z$, $\downarrow_z$ state associated to the original basis of the Hamiltonian. These results can be verified using spin-polarized transport at the edges and spin-polarized ARPES. The approach for the light-matter interaction resolved in the two Dirac valleys is also equivalent to a pair of one-half topological numbers associated to $\uparrow_z$ or $\downarrow_z$ spin polarizations. A pair of half topological invariants is also revealed from the equatorial plane on the sphere. This pair of half invariants then links with the physical properties of the topological semimetal on the bilayer model \cite{hutchinson_quantum_2021,OneHalfKLH}. We also mention here that one-half topological numbers also find applications for the physics of interacting topological superconducting p-wave wires \cite{OneHalfwires} and the physics of Majorana fermions \cite{OneHalfKLH,MajoranaSpheres}.

These protected topological semimetals present then two different flows of charges, associated to a two-dimensional Fermi liquid in the bulk and a one-dimensional (protected) channel(s) at the edges which may find applications in electronics and energy.


We acknowledge useful discussions with Joel Hutchinson, Andrej Mesaros and Michel Ferrero for interesting discussions. This work was supported by the french ANR BOCA and the Deutsche Forschungsgemeinschaft (DFG), German Research Foundation under Project No. 277974659. Sariah Al Saati thanks Ecole Polytechnique for funding his PhD thesis. 

\appendix
\section{Light-Matter Interaction and Topological Responses}
\label{lightmatter}

Here, we introduce the general definitions related to the light-matter coupling and remind the main steps towards the derivation of Eq. (\ref{quantizedresponse}) for the two-bands Haldane model.

When introducing the light-matter coupling on the honeycomb lattice, we can build very general conservation laws for the photo-induced current density (see precisely 
Sec. 3.4 of Ref. \cite{ReviewKLH} and \cite{PRBLight,KLHLight}) from the reciprocal space. Within this formulation, which is also equivalent to the dipole theory in Ref. \cite{Goldman}, the kinetic term in the Hamiltonian which does not commute with $\sigma_z$ will then produce the form of the current density. 
Within the Dirac approximation, we take the same definitions of the Brillouin zone as in \cite{PRBLight,ReviewKLH}, such that the (photo-induced) current operator takes the form,
\begin{equation}
\hat{J}_{\zeta}(t) = v_F(p_x+A_x(t))\sigma_y - (\zeta p_y + A_y(t))\sigma_x
\end{equation}
with the velocity $v_F=\frac{3}{2\hbar}ta$ and ${\bf p}=(p_x,p_y)$ measures a small {\it wave-vector} deviation from the Dirac points. In this way, we precisely introduce ${\bf k}={\bf K}+{\bf p}$ and similarly for the ${\bf K}'$ point.
Here, $\zeta=\pm 1$ at the $K$ and $K'$ respectively such that within the choice of the Brillouin zone the two Dirac points are related through $p_y\rightarrow -p_y$ or more generally $k_y\rightarrow -k_y$. The vector potential due to the classical light source is included through a shift of the wave-vector. It is important to emphasize that this form of the current density is also applicable for the two-bands Haldane model due to the fact that the charge density can be written in terms of the projection operators $\hat{P}_a = \frac{1}{2}(\mathbb{I}+\sigma_z)$ and $\hat{P}_b = \frac{1}{2}(\mathbb{I}-\sigma_z)$ on each sub-lattice \cite{ReviewKLH,PRBLight}. Through the Heisenberg equations of motion, the Haldane term $d_z$ then commutes with $\sigma_z$. This comes from the fact that inter-band transitions through light requires the pseudo-spin to flip at the Dirac points from the structure of eigenstates within the topological phase. Interestingly, the topological structure of the band theory in the Haldane model will naturally occur through the evaluation of the averaged photo-induced responses within the linear response theory. There is then a subtlety in the definitions to be careful with: $\zeta$ refers both to the direction of light polarization, right-handed (+) or left-handed (-), and to the Dirac point of interest the light will resonate with from energy conservation \cite{ReviewKLH,KLHLight}
\begin{eqnarray}
\label{vectorpotential}
\hbox{Re} A_x &=& A_0 \cos(\omega t) \\ \nonumber
\hbox{Re}A_y &=& -\zeta A_0  \sin(\omega t).
\end{eqnarray}
The light-matter Hamiltonian can be written in analogy to a spin-1/2 particle where $\sigma^+$ and $\sigma^-$ operators correspond to light-induced hopping terms from one to the other sublattice in the plane \cite{PRBLight,ReviewKLH,KLHLight}. 
Within the topological phase of the Haldane model, due to the energy conservation and inversion effects between the specific form of eigenstates at the two Dirac points \cite{ReviewKLH,KLHLight}, the right-handed circular polarization interacts with one Dirac point and the left-handed circular polarization interacts with the other Dirac point. From the form of the Dirac Hamiltonian for one layer we also have
\begin{eqnarray}
\sigma_x &=& \frac{1}{\hbar v_F}\frac{\partial {\mathcal H}}{\partial p_x} \\ \nonumber
\sigma_y &=& \frac{1}{\hbar v_F}\frac{\partial {\mathcal H}}{\partial (\zeta p_y)}.
\end{eqnarray}
\begin{equation}
\label{lightequation}
\hat{J}_{\zeta}(t) = (p_x+A_x(t))\frac{\partial {\mathcal H}}{\hbar\partial (\zeta p_y)} - (\zeta p_y + A_y(t))\frac{\partial {\mathcal H}}{\hbar\partial p_x}.
\end{equation}
Since we will evaluate the linear response to second-order in $A_0$, as if we just keep the terms in $A_x(t)$ and $A_y(t)$ in this formula, this is equivalent to measure the  light-induced response at the Dirac points with here $p_x=p_y=0$ or to average on all the wavevectors within the Brillouin zone \cite{Goldman}. Therefore, this results in the form of the current operator 
\begin{equation}
\label{Jzeta}
\hat{J}_{\zeta}(t) = \frac{1}{2\hbar}A_0 e^{-i\omega t} \left( \frac{\partial {\mathcal H}}{\partial (\zeta p_y)} +\zeta i  \frac{\partial {\mathcal H}}{\partial p_x}\right) +h.c.
\end{equation}
For the Haldane model, this leads to the following response associated to the {\textit{ photo-induced currents}} \cite{PRBLight,ReviewKLH}
\begin{eqnarray}
\label{zeta}
\Gamma_{\zeta}(\omega) &=& \frac{2\pi}{\hbar} \frac{A_0^2}{2} \left|\langle + | \left( \frac{\partial {\mathcal H}}{\hbar\partial (\zeta p_y)} +\zeta i  \frac{\partial {\mathcal H}}{\hbar\partial p_x}\right)| - \rangle\right|^2 \nonumber \\ 
&\times& \delta(E_l({\bf p})- E_u({\bf p}) - \hbar\omega).
\end{eqnarray}
The symbol $\zeta$ only occurs through $\zeta^2=1$ which means that it can be dropped at this step of the calculation. It is then important to remind here that we have the correspondences of eigenstates $|+\rangle_K \leftrightarrow |-\rangle_{K'}$ and $|-\rangle_K \leftrightarrow |+\rangle_{K'}$ for the Haldane model such that the crossed terms result in a finite number when measuring $\Gamma_+(\omega)-\Gamma_-(\omega)$. At this step the $\delta$-function is applicable close to both Dirac points; $E_l$ and $E_u$ refer to the lower and upper energy bands with energy $\pm |{\bf d}|$ in the Haldane model. These crossed terms are precisely related to the function entering into the quantum Hall response since we can introduce precisely the velocities $v^x$ and $v^y$. It is possible to find a resonance condition for one light polarization only setting $E_u(K)-E_l(K)=\hbar\omega$ with ${\bf p}\rightarrow {\bf 0}$. This implicitly assumes to have a precise knowledge of the energy spectrum from other measures. In this way, we can equivalently (mathematically) integrate on all frequencies since the $\delta$-function is zero when $E_u-E_l\neq \hbar\omega$. Then, this results in a quantized circular dichroic response related to the topological invariant of the lowest and upper bands $C=\mp 1$  \cite{Goldman,PRBLight,ReviewKLH,KLHLight}
\begin{equation}
\label{quantizedresponse}
\left| \frac{\Gamma_{+} - \Gamma_{-}}{2} \right| = \frac{2\pi}{\hbar} A_0^2 |C|,
\end{equation}
which can be then equivalently resolved from the information at the Dirac points. This topological quantization from the Dirac points was verified numerically and also in the presence of interactions through the stochastic variational approach that we developed \cite{PRBLight}.

 It is also relevant to mention that finding the resonance with the $M$ point(s) or (equivalently $\Gamma$-point) this would result in a half-quantized response such that this result can also be found from an exact result on the lattice \cite{KLHLight}. 

\section{Berry Functions and Light}
\label{Berry}

For the evaluation of the Berry curvature $F_{p_x p_y}$ in Eq. (\ref{lightberry}), associated to the responses to circularly polarized light, we proceed similarly as in Ref. \cite{KLHLight} and generalize this equation for the different bands of the topological semimetal. 
We can then evaluate the Berry curvatures from the definition around each Dirac point
\begin{equation}
F_{p_x p_y} = i\frac{\langle +|\partial_{p_x} \mathcal{H}|-\rangle \langle -| \partial_{p_y}\mathcal{H} |+\rangle - \langle +|\partial_{p_y} \mathcal{H}|-\rangle \langle -|\partial_{p_x} \mathcal{H} |+\rangle}{(E_- - E_+)^2}.
\end{equation}
We equally have
\begin{equation}
\label{functions}
F_{p_x p_y} = i(\hbar v_F)^2\zeta \frac{\langle +| \sigma_x |-\rangle \langle -|\sigma_y|+\rangle - \langle +|\sigma_y |-\rangle \langle -| \sigma_x|+\rangle}{(E_- - E_+)^2}.
\end{equation}
For a pair of bands associated to the Haldane model, $E_u=E_+$ and $E_-=E_l$. With the eigenstates on the sphere this results in \cite{KLHLight}
\begin{equation}
F_{p_x p_y} = \zeta\frac{\hbar^2 v_F^2}{2|{\bf d}|^2}\cos\theta_{{\bf p}}.
\end{equation}
See also a similar equation associated to the Dirac equation and Thomas procession \cite{ShankarMathur} if we identify $\langle \sigma_z\rangle=\cos\theta_{\bf p}$ from Ehrenfest theorem. This definition reproduces precisely the quantum Hall conductivity of the ground state in the Haldane model from the Dirac approximation of the kinetic term \cite{ReviewKLH} and this form of $F_{p_x p_y}$ can be verified when developing the eigenstates close to the Dirac points \cite{ReviewKLH,Ryu}. For this evaluation, we identify the correspondence between lattice and sphere through the form
of the ${\bf d}_{\bf k}$ vector with the Dirac approximation of the kinetic term
\begin{equation}
|{\bf d}_{\bf k}| (\cos\varphi\sin\theta,\sin\varphi\sin\theta,\cos\theta) = (\hbar v_F|{\bf p}| \cos\tilde{\varphi}, \hbar v_F |{\bf p}|\sin(\zeta \tilde{\varphi}),\zeta m)
\end{equation}
where $m=3\sqrt{3}t_2$ in the Haldane model. In the vicinity of the Dirac points we observe that 
\begin{eqnarray}
p_x &=& \frac{|{\bf d}_{\bf k}|}{\hbar v_F}\cos\varphi\sin\theta  \nonumber \\
p_y &=& \zeta\frac{|{\bf d}_{\bf k}|}{\hbar v_F}\sin\varphi\sin\theta.
\end{eqnarray}
In this case e.g. close to $\theta=0$, $\tan\theta\sim \theta=\frac{\hbar v_F}{m}|{\bf p}|$ with $\varphi=\tilde{\varphi}$. This choice of representation is especially judicious to describe the effect of circular motion locally around each Dirac point in the plane.

It is interesting to observe that the topological invariant can be measured from these Berry curvatures locally at the Dirac points \cite{KLHLight}
\begin{equation}
\label{Fresponses}
F_{p_x p_y}(K) - F_{p_x p_y}(K') = C\frac{(\hbar v_F)^2}{m^2}.
\end{equation}
The prefactor $\frac{(\hbar v_F)^2}{m^2}$ comes from the fact that the sphere precisely corresponds to identify $\hbar \frac{v_F}{a}=m$ and when $\frac{\hbar v_F}{a}\neq m$ then we have an ellipse. Measuring the addition of the two
Berry curvatures on the lattice locally at the Dirac points also reproduces the quantum Hall response within this map, justifying also why we can also measure the responses to circularly polarized light at the Dirac points.
It is interesting to comment that Eq. (\ref{Fresponses}) is also applicable to the lowest band of the energy spectrum due to the particular form of the Hamiltonian in Eq. (\ref{eq:Hmodel}) and due to the fact that only bands with the same spin number
can enter in the definition of $F_{p_x p_y}$.

It is also possible to evaluate the transition rates or inter-band transition probabilities in the presence of a circular polarization
\begin{equation}
\delta \mathcal{H}_{\pm} = A_0 e^{\pm i \omega t} \sigma^+ +h.c. 
\end{equation}
These transition rates are measurable through the variation in time of the population in a given band associated to the ground state (whereas the difference of photo-induced currents related to circular polarizations would give zero).
For the Haldane two-bands model, they take the form \cite{KLHLight}
\begin{eqnarray}
\Gamma_{\pm}(\omega) = \frac{2\pi}{\hbar}A_0^2 \alpha(\theta) \delta(E_b-E_a\mp \hbar\omega),
\end{eqnarray}
where
\begin{eqnarray}
\alpha(\theta) = \cos^4\frac{\theta}{2} +\sin^4\frac{\theta}{2}.
\end{eqnarray}
The resonance will be found with the right-handed wave when $E_b-E_a=E_+-E_-=+\hbar \omega$ and the resonance will be found with the left-handed wave when $E_b-E_a=E_- - E_+=-\hbar\omega$ such that
either $\sigma^+$ or $\sigma^-$ contributes for a specific light polarization. For $\theta=\frac{\pi}{2}$ we obtain the precise identity  \cite{KLHLight}
\begin{equation}
\langle -| \sigma_x |+\rangle \langle +| \sigma_x |-\rangle = \frac{C^2}{2}
\end{equation}
which corresponds to a linear polarization as a result of the superposition of two circular polarizations. When $\theta=\frac{\pi}{2}$ this measures precisely a $\frac{1}{2}$ response compared to the situation at the $K$ Dirac point. We can generalize this result for the topological semimetal as follows. The light-induced transition between the lowest energy band described by the eigenstate $|\downarrow_x\rangle\otimes|-\rangle$ and the (third) energy eigenstate described by $|\downarrow_x\rangle\otimes|+\rangle$ reveals precisely the same $\frac{1}{2}$ response when $\theta=\theta_c^-$. Similarly, the light-induced transition between the second and the fourth energy bands corresponding to eigenstates $|\uparrow_x\rangle\otimes|-\rangle$ and $|\uparrow_x\rangle\otimes|+\rangle$ reveals the same $\frac{1}{2}$ response. When $\theta=\theta_c^+$, there is an inversion between the two middle bands such that the system cannot couple resonantly to light revealing the formation of the nodal ring semimetal. Therefore, measuring a pair of half responses at $\theta_c^-=\frac{\pi}{2}-\epsilon$ with $\epsilon\rightarrow 0$ is also equivalent to measure a quantized quantum Hall response $\frac{e^2}{h}$ similar to a half-Skyrmion.

\bibliographystyle{crunsrt}
\bibliography{biblioPaper}

\end{document}